\documentclass[aps,physrev,twocolumn,groupedaddress,floatfix]{revtex4-2}
\usepackage{amsmath}
\usepackage{amsfonts}
\usepackage{graphicx}
\usepackage{dcolumn}
\usepackage{float}

\begin{document}


\title{\textbf{Examining the evolution of phase-space elements for \textit{C.elegans} locomotion}}


\author{Dimitrios Tzepos}
\author{Jenny Magnes}
\email[]{Contact author: \url{jemagnes@vassar.edu}}
\affiliation{Vassar College, Department of Physics \& Astronomy\\
Poughkeepsie, NY, United States of America}


\date{\today}

\begin{abstract}
The \textit{Caenorhabditis Elegans} (\textit{C.elegans}) nematodes have long been a model organism for quantitative behavioral analysis, due to their tractable nervous system and well-characterized genetics. In particular, dynamic diffraction has been a successful method of studying said microorganisms due to its low level of noise and the ability to simultaneously study multiple degrees of freedom of their neuromuscular system through their locomotion. In this study, we estimate the Lyapunov spectrum of \textit{C.elegans} locomotion, which offers an insight into how volume elements evolve in the phase space of the underlying dynamical system. For that, we used the Sano-Sawada algorithm to estimate the spectra from the trajectories reconstructed using the Takens embedding procedure. In total, two positive and one negative exponents were calculated and verified to be non-spurious through investigations of their stability for different sets of parameters. Those exponents have values of $0.860 \pm 0.028$, $0.389 \pm 0.014$, and $-3.451 \pm 0.074$ respectively. The presence of two positive exponents indicates that \textit{C.elegans} locomotion is hyperchaotic, while the total sum being negative indicates that the system is dissipative and non-Hamiltonian. Those are key observations for the underlying system and will be significant for the potential creation of future mathematical or computational models.
\end{abstract}


\maketitle

\section{Introduction}

The study of animal locomotion has long served as a fertile proving ground for nonlinear dynamical methods in physics and applied mathematics.  
Due to its tractable nervous system and well-characterized genetics, the nematode \textit{Caenorhabditis elegans} (\textit{C.elegans}) occupies a privileged position among model organisms for quantitative behavioral analysis \cite{pierce2008,gray2005}.  
High-speed imaging now yields time series of postural and centroid coordinates with millisecond resolution, inviting a phase-space treatment of the worm’s motion that goes beyond purely kinematic descriptions.

Previous work has demonstrated that \textit{C.elegans} locomotion displays markers for deterministic chaos \cite{jenny}. This work investigates whether the instability extends to more than one phase-space direction, i.e., hyperchaos \cite{rossler}.  

Although a positive Largest Lyapunov Exponent (LLE) signals a sensitive dependence on initial conditions \cite{wolf,rosenstein}, the full Lyapunov spectrum is required to (i) diagnose hyperchaos, defined by two or more positive exponents \cite{rossler} and (ii) evaluate the sum of exponents, whose sign distinguishes Hamiltonian from dissipative flow \cite{ottbook}. 

Extracting a spectrum from experimental data is nontrivial: measurements can be noisy, and the underlying dynamics are unknown and available only through scalar time series. We therefore employ Takens’ embedding theorem \cite{takens} to reconstruct the phase space and the Sano–Sawada algorithm \cite{sano} to estimate local Jacobians, allowing direct computation of Lyapunov exponents from the embedded trajectory.  
Robustness is verified by sweeping both the embedding dimension and the neighborhood radius, following the practices outlined in previous investigations \cite{kantz1997,skokos2016}.  

Our main result is the observation of two statistically significant positive exponents across nine independent datasets of freely swimming worms, complemented by a strongly negative exponent that drives phase-space contraction.  

The spectrum thus indicates hyperchaotic behavior and, by virtue of the sum of Lyapunov exponents being negative, dissipative behavior.  
These findings extend earlier reports of just the LLE of \textit{C. elegans} locomotion \cite{jenny,tzepos} and provide quantitative benchmarks for future theoretical and computational models.

The remainder of the paper is organized as follows: Section II summarizes the experimental system and the data acquisition protocol. Section III reviews the theoretical background and describes the embedding procedure. Section IV describes the Sano-Sawada algorithm. Section V presents the Lyapunov spectrum estimates and assesses their sensitivity to analysis parameters. Section VI provides a discussion of the results' implications. Section VII concludes the findings.

\section{Physical System}

The experimental system described by Magnes et al. \cite{jenny} consists of freely swimming adult wildtype \textit{C.elegans} hermaphrodites in a water-filled optical cuvette under conditions that minimize external stimuli. Each worm is imaged using a dynamic optical diffraction setup, where a coherent light source oversamples the specimen, and the resulting far-field diffraction pattern is recorded over time at one point in the diffraction pattern. This optical configuration captures global postural information without the need for invasive tracking or body segmentation, allowing for the high-throughput acquisition of behavioral time series.

Diffraction patterns are collected using a photodiode (PD) at a rate of $1.1$ kHz with a spatial resolution sufficient to resolve the temporal evolution of locomotor dynamics. Each data set is 40-60 seconds in length, producing scalar-valued time series corresponding to the diffraction intensity integrated over a defined detector region. The resulting signal encodes the aggregate motion of the worm and serves as the observable from which the underlying dynamics is inferred.

All nematodes are synchronized in age and kept under standard growth conditions prior to the sampling of their locomotory dynamics. To reduce variability due to environmental effects, all recordings are performed at a constant temperature and under uniform lighting conditions. Worms are selected for analysis based on sustained forward locomotion, avoiding periods of reversal or inactivity.

In this framework, we interpret the measured time series as a projection of the full phase-space trajectory of a higher-dimensional dynamical system. The assumption is that the behavior arises from a smooth deterministic process that governs the neuromuscular activity of the worm and that this process can be meaningfully reconstructed from the observable scalar using the Takens embedding method. This reconstructed trajectory forms the basis for all subsequent Lyapunov-spectrum estimation.

\section{Theory}
\subsection{Lyapunov Spectra}

Consider a differentiable function $f:\mathbb{R}^d \to \mathbb{R}^d$ such that, for a certain discrete-time dynamical system, its state is given by the map:
\begin{equation}
    \mathbf{x}_{n+1}=f(\mathbf{x}_{n}) \textrm{  ,  }n=1,2,3,...
\end{equation}
which is defined on a $d$-dimensional phase space such that $\mathbf{x}_{n},\mathbf{x}_{n+1}$ are vectors with $d$ elements. For our purposes, this represents the underlying dynamical system that governs the neuromuscular activity of the \textit{C.elegans} after being projected onto a discrete time series and re-embedded into a phase space of an appropriate dimension $d$ (see more at III-C).

The corresponding tangent map, which describes the evolution of vectors in the tangent space of the system's trajectories, is given by:
\begin{equation} \label{jacobian}
    \mathbf{w}_{n+1} = \mathbf{T}(\mathbf{x}_n)\mathbf{w}_n \textrm{ , } \mathbf{T}_{ij}(\mathbf{x}_n) = \frac{\partial f_i}{\partial x_j}\bigg\rvert _{\mathbf{x}=\mathbf{x}_n}
\end{equation}
where $\mathbf{T}(\mathbf{x}_n)$ is the Jacobian operator of the map defined at that point. For some initial tangent vector $\mathbf{w}_0$:
\begin{equation} \label{T^n}
    \mathbf{w}_{n}= \mathbf{T}(\mathbf{x}_{n-1})\mathbf{T}(\mathbf{x}_{n-2})...\mathbf{T}(\mathbf{x}_1)\mathbf{w}_0 = \mathbf{T}^n\mathbf{w}_0
\end{equation}
where $\mathbf{T}^n$ is a linear operator \cite{skokos}. The matrix $\mathbf{\Lambda}$ is defined by:
\begin{equation}
    \mathbf{\Lambda} = \lim_{n\to \infty}\frac{1}{n}\ln(\mathbf{T}^n)
\end{equation}
and its $d$ ordered eigenvalues $\lambda_1 \geq \lambda_2 \geq ... \geq \lambda_d$ are called the Lyapunov spectrum of characteristic exponents of the system, or simply its Lyapunov spectrum \cite{vastano}.

The existence of the limit that defines $\mathbf{\Lambda}$ is ensured by the multiplicative theorem of Osedelec \cite{ergodic}, which requires the existence of an ergodic invariant probability measure. In qualitative terms, for $n \to \infty$, a trajectory should pass through every point in phase space. A system with such a measure is called ergodic, and its Lyapunov spectrum is invariant for all initial conditions.

Traditionally, a positive Largest Lyapunov Exponent ($\lambda_1$) is associated with deterministic chaos. If a system has a positive $\lambda_1$, then the vectors in the tangent space of a trajectory will diverge exponentially over time. Classical chaos is defined by the divergence of nearby trajectories \cite{strogatz}, and henceforth a positive $\lambda_1$ indicates that said system is most likely chaotic. Due to the ease with which it can be calculated even in the presence of noise \cite{jenny}, it is a very popular method for investigating the presence of chaos in experimental systems.

The presence of two or more positive Lyapunov exponents in a dynamical system is the criterion for it to be considered hyperchaotic, a term first coined by Rossler \cite{rossler}. Though calculating $\lambda_2$ is significantly more challenging due to $\lambda_1$ frequently dominating computations \cite{wolf}, numerous experimental detections of hyperchaos have been made \cite{chem,nmr}.

In the context of \textit{C.elegans} locomotion, $\mathbf{x}_n$ represents the state of the underlying system for some discrete iteration $n$. Hence, the detection of chaos or hyperchaos indicates that the locomotion of the nematodes, though deterministic, is inherently unpredictable, since any initial uncertainty in the state of the underlying dynamical system will grow exponentially. Although showing that the system is hyperchaotic does not have any immediate qualitative implications, estimating its Lyapunov spectrum gives us a better idea of how fast uncertainties in the initial conditions grow within the system. Specifically, this rate is given by the sum of the positive Lyapunov exponents of the system, as it also represents the rate of creation of information over time \cite{strogatz}.

\subsection{Liouville's Theorem and Dissipativity}
The Lyapunov exponents of a dynamical system describe the exponential growth of vectors along directions tangent to phase-space trajectories. The components of the vectors along the directions associated with positive exponents grow, while those associated with negative ones approach 0 in magnitude. We can therefore approximate the average growth rate of the $k$-dimensional volume elements in phase space given the sum of the first $k$ Lyapunov exponents\cite{ergodic}.

A corollary of Liouville's Theorem states that if the sum of Lyapunov exponents of a system is 0:
\begin{equation}
    \sum_{i=1}^{d}\lambda_i = 0,
\end{equation}
then the system is Hamiltonian, since the sum being zero means that phase-space volumes are preserved. By definition, the system's dynamics can be expressed in its Hamiltonian, and if that is time-independent, then the system also conserves energy \cite{scholarpedia}.

If the sum of the exponents is not zero, the system is non-Hamiltonian \cite{shivamogi}. Those systems are called dissipative and do not conserve their energy due to the presence of some external factor (friction, resistance, etc.). Dissipative systems are also considered to be thermodynamically open, which means that they exchange energy with their surroundings. Additionally, there exists no Hamiltonian function that encompasses their dynamics.

\subsection{Takens' Embedding Theorem and Spurious Exponents}
To investigate \textit{C.elegans} movement, we assume the presence of a deterministic underlying dynamical system, and the time series obtained from dynamic diffraction is considered to contain information from all degrees of freedom of said system \cite{jenny}. Takens embedding is a reliable method from which the original dynamics can be recovered \cite{takens}. 

\begin{center}
\begin{figure}
\includegraphics[width=0.45\textwidth]{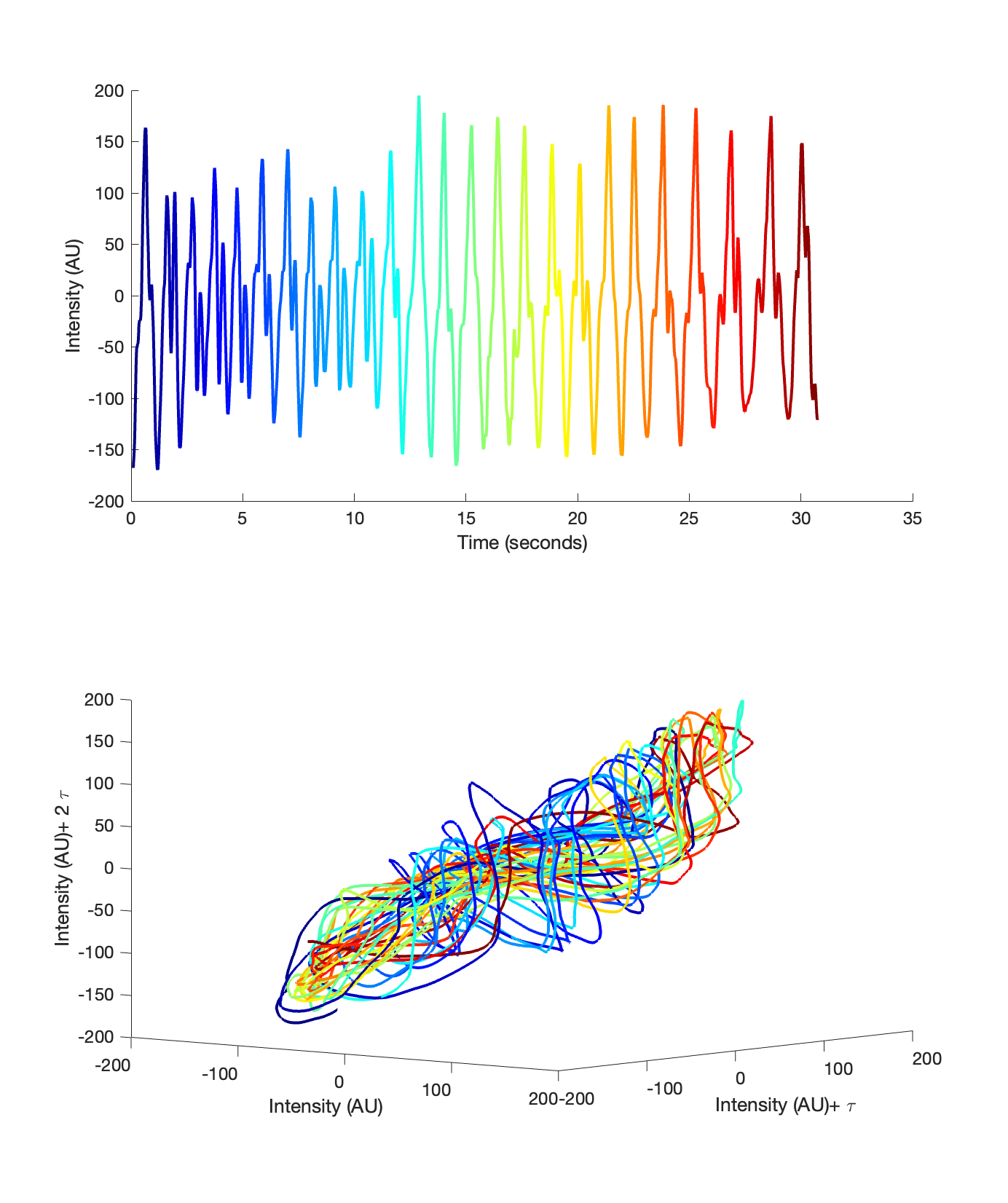}
\caption{\label{fig:dataset} Lag plot (bottom) of a time series (top) used in the analysis below, visualized in a projected 3-dimensional phase space using the Takens procedure. The changing color represents evolution in time. Selecting $d=3$ seems to resolve most of the attractor, but according to the results from the FNN percentage \cite{jenny}, a $d$ of at least 4 is required to properly resolve the trajectories. The axes of the figure are unmarked because they are in arbitrary units (AU), just like the pre-embedding time series.}
\end{figure}
\end{center}

Consider a time series $x_T = \{x_1,x_2,...,x_N\}$ of length $N$. We can embed this time series into a $d$-dimensional phase space through the transformation:

\begin{equation} \label{emb}
    x_T' = \left\lbrace \begin{pmatrix}
x_1 \\
x_{1+\tau} \\
\vdots \\
x_{1+(d-1)\tau} 
\end{pmatrix},\begin{pmatrix}
x_2 \\
x_{2+\tau} \\
\vdots \\
x_{2+(d-1)\tau} 
\end{pmatrix},..,\begin{pmatrix}
x_{N-(d-1)\tau} \\
x_{N-(d-2)\tau} \\
\vdots \\
x_{N} 
\end{pmatrix} \right\rbrace,
\end{equation}

where $d$ is called the embedding dimension and $\tau$ is the lag. The embedding dimension is the dimension of the phase space in which we choose to reconstruct the trajectory, and the lag is the time delay between two coordinates of the said space (see Eq.\ref{emb}). According to Takens' embedding theorem, the function that maps the phase-space trajectory of the original system to the embedded trajectory is a diffeomorphism \cite{takens}. The Lyapunov spectrum is an invariant quantity under diffeomorphic transformations \cite{ergodic}, so if we estimate it for the reconstructed trajectory, we recover the exponents of the original system. An example of one of the \textit{C.elegans} time series embedded with $d=3$ can be seen in Fig.\ref{fig:dataset}.

The choice of parameters $d$ and $\tau$ that best resolve a trajectory can be arbitrary, but there exist methods to determine the most suitable values. An adequate lag can be estimated through the first minimum in Mutual Information \cite{mi}, while a proper embedding dimension is the lowest one in which no or very few False Nearest Neighbors are observed \cite{fnn}. The optimal lag of our datasets was calculated in a previous publication \cite{jenny}

The embedding method might introduce spurious exponents \cite{tepkin} when working with Lyapunov spectra. If one attempts to embed a time series at a higher dimension than that of the original system, they will recover the original exponents and also a few extra ones that do not belong to the system. Those are called spurious exponents, and they are artifacts of the embedding process. It is generally difficult to know whether an exponent is spurious or not. Although attempts have been made \cite{time_reversal}, there is no reliable method to identify spurious exponents. It has been observed that spurious exponents are most often negative and also that their values are less stable when the embedding dimension is varied \cite{chem}. Therefore, it suffices to estimate the spectra for a range of embedding dimensions and look for exponents that appear even when $d$ is small, and are also more stable as we increase it.

\section{The Sano-Sawada Algorithm}
The Sano-Sawada algorithm \cite{sano} is used to estimate Lyapunov spectra from their reconstructed trajectories. For this work, we used the TISEAN library \cite{tisean} for nonlinear time series analysis.

\begin{center}

\begin{figure}
\includegraphics[width=0.45\textwidth]{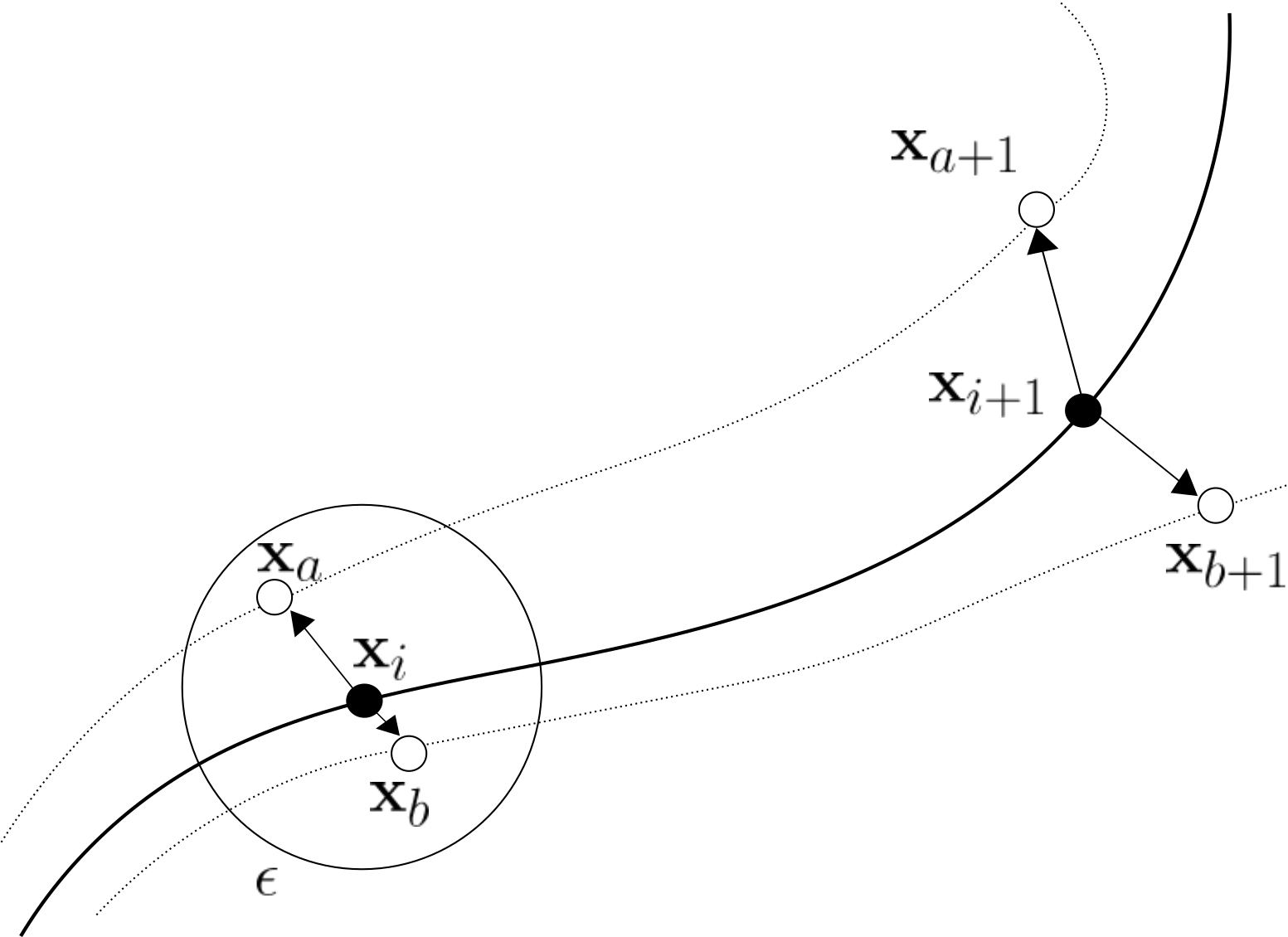}
\caption{\label{fig:sano-sawada}The process of selecting a neighborhood of radius $\epsilon$ around a particular point $\mathbf{x}_i$ on the reconstructed trajectory. The points $\mathbf{x}_a,\mathbf{x}_b$ within this neighborhood are considered neighboring points. Once an iteration has been completed, these points are mapped to $\mathbf{x}_{a+1},\mathbf{x}_{b+1}$ respectively. By analyzing the initial and final separation vectors from the trajectory point to its neighbors, we can identify the linear operator $A_j$ that corresponds to each neighboring point.}
\end{figure}
\end{center}

For each point $\mathbf{x}_i$ on the discrete reconstructed trajectory, consider a small sphere of radius $\epsilon$ around the point and the set of all other trajectory points $\mathbf{x}_j$ inside the sphere (Fig.\ref{fig:sano-sawada}). From now on, we shall refer to this set as the neighborhood of $\mathbf{x}_i$ and to the points in it as the neighboring points of $\mathbf{x}_i$. Consider now the set of vectors:
\begin{equation} \label{sep}
    \{\mathbf{y}_j\} = \{\mathbf{x}_j-\mathbf{x}_i \textrm{ }\textrm{ such that }\textrm{ }|\mathbf{x}_j-\mathbf{x}_i|<\epsilon\}.
\end{equation}
In the next iteration, $\mathbf{x}_i$ is mapped to $\mathbf{x}_{i+1}$ and all neighboring points $\mathbf{x}_j$ will be mapped to $\mathbf{x}_{j+1}$. Therefore, each vector $\mathbf{y}_j$ will be mapped to some $\mathbf{z}_j$ such that:
\begin{equation} \label{sep}
    \{\mathbf{z}_j\} = \{\mathbf{x}_{j+1}-\mathbf{x}_{i+1} \textrm{ }\textrm{ such that }\textrm{ }|\mathbf{x}_j-\mathbf{x}_i|<\epsilon\}.
\end{equation}
For each, there exists a linear operator $A_j$ such that:
\begin{equation}
    \mathbf{z}_j = A_j \mathbf{y}_j.
\end{equation}
Assuming that $\epsilon$ is sufficiently small, $\{\mathbf{y}_j\}$ and $\{\mathbf{z}_j\}$ can be considered good approximations of the vectors in the tangent space of the trajectory, and performing a least-squares fit on the set of operators $A_j$ for some $\mathbf{x}_i$ yields $A^i$, which is an approximation of the local Jacobian $\mathbf{T}(\mathbf{x}_i)$ defined in Eq.\ref{jacobian}. For a time series of length $N$ and framerate $1/t$, the Lyapunov exponents $\lambda_i$ can be estimated from:
\begin{equation} \label{lambda}
    \lambda_i = \lim_{N\to \infty}\frac{1}{Nt}\sum_{j=1}^{N}\ln |A^j e_j^i|
\end{equation}
where $\{e_j^i\}(i=1,2,...,d)$ is a set of basis vectors for the tangent space of the trajectory, determined by Gram-Schmidt orthonormalization \cite{gram}.

It follows from the above description that the neighborhood size $\epsilon$ significantly affects the algorithm's estimates. If it is too small, there are not enough neighboring points available to produce a reliable average from Eq.\ref{lambda}. If it is too large, the separation vectors defined in Eq.\ref{sep} cannot approximate the tangent space. Hence, the optimal way to ensure that the estimates are reliable is to calculate the Lyapunov spectra for different $\epsilon$, and verify that the estimates are robust for a significant range of $\epsilon$ values \cite{greek, chem}.  

\section{Results}

\subsection{Variation of estimates with embedding dimension}

The 10 datasets analyzed were consistent in terms of the magnitudes of the estimated exponents, as well as the locations where spurious exponents formed as the embedding dimension was increased.

\begin{figure}
\includegraphics[width=0.48\textwidth]{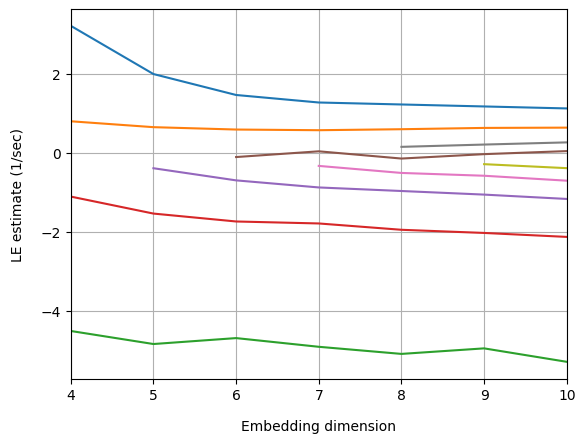}
\caption{\label{fig:embed} Sample graph for one dataset. The different colored lines stand for different exponents. New exponents form as $d$ is increased.}
\end{figure}

Starting as early as embedding dimension 4, two positive exponents and one negative are present for every dataset. As the dimension increases, more negative exponents form, which are of a smaller magnitude than the most negative one. Those exponents do not stabilize for some range of embedding dimensions, which is in accordance with the behavior of spurious exponents exhibited in previous studies that performed this analysis \cite{chem}. Hence, we can assume that these may be spurious. The same holds for what looks like a zero exponent, which forms for dimension 6, as well as a spurious positive exponent that shows when the embedding dimension is 8 or higher.

For the purposes of this study, we are interested in estimating only the two largest positive exponents and the most negative one. Therefore, we seek some dimension at which their values have stabilized. As shown in Fig.\ref{fig:embed}, and in agreement with the rest of our datasets, this embedding dimension is 7. Consequently, this value of $d$ is used to embed the time series for subsequent analysis.

\subsection{Variation of estimates with neighborhood size}

\begin{figure}
\includegraphics[width=0.48\textwidth]{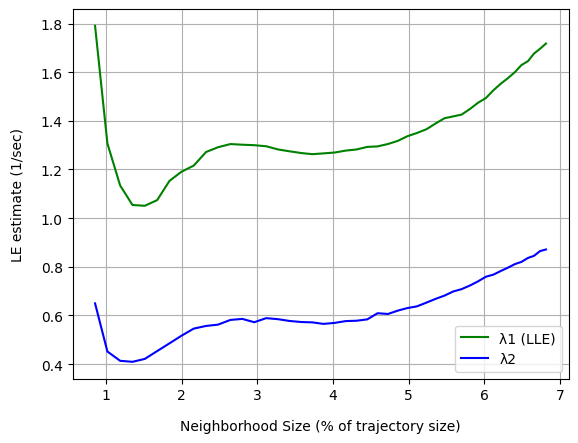}
\caption{\label{fig:k_graph} The estimates of $\lambda_1$ and $\lambda_2$ for a sample time series are plotted for a range of different values of $\epsilon$. For both exponents, a plateau is visibly formed in the middle of the graph, corresponding to a region where the estimates can be considered robust.}
\end{figure}

The estimates provided by the Sano-Sawada algorithm can be different depending on the value of the neighborhood size $\epsilon$ that we choose, and therefore the method can only be trusted if the estimates are robust for a significant range of values of $\epsilon$. To test that, the exponents are calculated for a range of $\epsilon$ values, and the estimates are graphed, as in Fig.\ref{fig:k_graph}. The plateaus observed for both exponents in the figure correspond to regions in the parameter space of the algorithm where the estimated values of the exponents are robust and can therefore be trusted.

\begin{figure}
\includegraphics[width=0.48\textwidth]{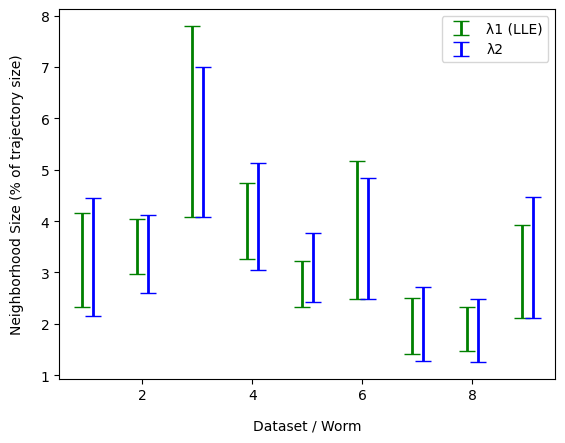}
\caption{\label{fig:plateaus} Range of $\epsilon$ values for which the graphs of each dataset's positive exponents plateau. Most of the Lyapunov exponents plateau in similar regions, at around $2-5\%$ of their respective trajectories.}
\end{figure}

In total, 9 out of the 10 analyzed datasets yielded plateaus in their respective graphs, and therefore, we see that the Sano-Sawada algorithm can be considered reliable for our purposes. The ranges of $\epsilon$ where the plateaus were detected for each dataset can be seen in more detail in Fig.\ref{fig:plateaus}. In both figures \ref{fig:k_graph} and \ref{fig:plateaus}, the values of $\epsilon$ have been represented as percentages of the total horizontal extent of each dataset's trajectory for comparison purposes. In this study, we define that as the maximum Euclidean distance between any two points in the trajectory\cite{chem}.

\subsection{Lyapunov Spectra Estimates}

For each of the 9 datasets that produced plateaus such as the one in Fig.\ref{fig:k_graph}, the values of $\lambda_1$ and $\lambda_2$ were estimated as the means of all the points in their respective plateaus. The individual error associated with each estimate represents the standard deviation of the mean of the set of points. The estimated values of all exponents, as well as the means of all datasets, are shown in Table \ref{table}.

\begin{table*}[hbt!]
\caption{\label{table} The $\lambda_1,\lambda_2,\lambda_7,\lambda_1+\lambda_2+\lambda_7$ estimates of all 9 datasets, including their individual error bars, the mean value of each exponent and its associated error given by the Standard Deviation of the Mean (SDM).}
\begin{ruledtabular}
\begin{tabular}{ccccc}
\textrm{Dataset / Worm}&
\textrm{$\lambda_1$ ($sec^{-1}$)}&
\textrm{$\lambda_2$ ($sec^{-1}$)}&
\textrm{$\lambda_7$ ($sec^{-1}$)}&
\textrm{$\lambda_1+\lambda_2+\lambda_7$ ($sec^{-1}$)}\\
\colrule
1 & $1.282\pm0.014$ & $0.573\pm0.011$ & $-4.843\pm0.051$ & $-2.988\pm0.076$\\
2 & $1.025\pm0.016$ & $0.396\pm0.054$ & $-4.104\pm0.031$ & $-2.683\pm0.101$\\
3 & $1.187\pm0.052$ & $0.577\pm0.027$ & $-4.003\pm0.016$ & $-2.239\pm0.095$\\
4 & $0.903\pm0.014$ & $0.422\pm0.013$ & $-3.776\pm0.028$ & $-2.451\pm0.055$\\
5 & $0.49\pm0.013$ & $0.233\pm0.009$ & $-2.75\pm0.015$ & $-2.027\pm0.037$\\
6 & $0.81\pm0.038$ & $0.372\pm0.023$ & $-3.026\pm0.016$ & $-1.844\pm0.077$\\
7 & $0.592\pm0.004$ & $0.252\pm0.004$ & $-2.973\pm0.009$ & $-2.129\pm0.017$\\
8 & $0.606\pm0.009$ & $0.253\pm0.006$ & $-2.839\pm0.022$ & $-1.98\pm0.037$\\
9 & $0.847\pm0.041$ & $0.423\pm0.022$ & $-3.559\pm0.015$ & $-2.289\pm0.078$\\
\textbf{Mean ($\pm$ SDM)} & \textbf{0.860 $\pm$ 0.028} & \textbf{0.389 $\pm$ 0.014} & \textbf{-3.451 $\pm$ 0.074} & \textbf{-2.292 $\pm$ 0.038}\\
\end{tabular}
\end{ruledtabular}
\end{table*}

In addition to the positive exponents, the least Lyapunov exponent $\lambda_7$ also exhibited remarkable consistency in its estimates, with similar plateaus. The corresponding results can be seen in Table \ref{table}, along with the sum $\lambda_1+\lambda_2+\lambda_7$, which can be considered an upper boundary for the sum of Lyapunov exponents (see more in Subsection VI-B).

\section{Discussion}
\subsection{Lyapunov Exponents and Hyperchaos}
According to the previous section, the estimated exponents for $\lambda_1$ and $\lambda_2$ remain significantly positive within their error margins. Their magnitudes are consistent and do not show significant deviations. In addition, the plateaus of the graphs, such as the one in Fig.\ref{fig:k_graph}, were clear for every dataset, as evidenced by the magnitude of the error bars for each dataset. This indicates that \textit{C.elegans} locomotion satisfies this criterion for hyperchaos.

Both $\lambda_1$ and $\lambda_2$ must be non-spurious to support the condition for hyperchaos. By varying the embedding dimension, the positive Lyapunov exponents form as early as embedding dimension 4. Spurious exponents do not appear unless the embedding dimension is larger than that of the underlying system \cite{tepkin}. Hence, assuming that this system is 4-dimensional or higher, we can be certain that $\lambda_1$ and $\lambda_2$ are non-spurious \cite{chem}. This is valid because the percentage of False Nearest Neighbors \cite{fnn} for \textit{C.elegans} locomotion becomes steady at embedding dimension 4, indicating that this is the minimum embedding dimension required to properly resolve a trajectory \cite{jenny}.

It is important to compare the values obtained here with those of our previous work \cite{jenny,tzepos}, which uses the Rosenstein \cite{rosenstein} and Wolf \cite{wolf} algorithms, respectively. Those works yielded an average LLE of $1.25\pm0.05 \textrm{ }sec^{-1}$ for \textit{C.elegans} locomotion. Based on the comparison, the Sano-Sawada algorithm might tend to underestimate the exponent magnitudes. However, a deviation of this size is expected, since the two previous methods exploit the geometric properties of the LLE, whereas the Sano-Sawada algorithm uses the Jacobian definition of the Lyapunov spectrum. Hence, since the estimates for both are positive and of similar magnitudes, it can be said that the three methods are consistent.

Classical chaos in a dynamical system is associated with lateral instability across some direction tangent to a given phase-space trajectory \cite{rossler}. Correspondingly, since hyperchaos is defined by the existence of two positive characteristic exponents, it manifests itself as instability across some tangent plane (defined by the two directions corresponding to the positive exponents).

In the introduction, it was highlighted that the primary aim of this study was to explore specific aspects of \textit{C. elegans} movement. This exploration aids in more accurately defining prospective models intended for describing the nematodes' neuronal networks. Calculating the positive characteristic exponents of a system is a relatively straightforward check. Therefore, a comparison can be drawn to determine whether a future model is accurate since we have estimated the Lyapunov exponents for the physical system.

\subsection{Dissipativity}
In addition to the positive exponents $\lambda_1$ and $\lambda_2$, the value of the least exponent $\lambda_7$ was also estimated. That is because those were the three exponents that form for an embedding dimension as low as 4, remain consistent as $d$ increases, and for which plateaus like the ones in Fig.\ref{fig:plateaus} were obtained. Hence, it can be reasonably assumed that those exponents are non-spurious.

Assuming that there are no more positive exponents in the Lyapunov spectrum of \textit{C. elegans} locomotion, the sum $\lambda_1+\lambda_2+\lambda_7$ can be considered an upper boundary for the sum of all Lyapunov exponents. The estimates for this sum (see Table \ref{table}) are significantly negative for all datasets, so it can be inferred that the sum of Lyapunov exponents for our system is negative.

According to the theory developed in section III-B, this suggests that the underlying dynamical system that we examine is non-Hamiltonian. This has two major implications. First, it tells us that the system's energy is not conserved over time. Hence, when modelling it, a friction-like term must be included to compensate for that. Secondly, we now know that the dynamics of it cannot be expressed with a Hamiltonian function. This is a significant finding because it can be useful when designing potential models of the \textit{C. elegans} neuronal circuit, much like our estimates for the Lyapunov exponents themselves.

\section{Conclusion}
This investigation has estimated the two positive Lyapunov exponents of \textit{C. elegans} locomotion using the Sano-Sawada algorithm. To verify the robustness of the calculated exponents and confirm they are genuine, variations in the embedding dimension and the neighborhood size $\epsilon$—a critical algorithm parameter—were applied, and the exponents were estimated within the $\epsilon$ ranges where their estimates remain stable. The values of the two exponents for all our data sets that yielded plateaus were strictly positive, indicating the presence of hyperchaos in the dynamics of the system. In addition, assuming that there are no more positive exponents in the system, we have shown that it behaves dissipatively, not conserving its energy, due to the presence of a negative least exponent, which was estimated in a similar manner as the two positive ones. 

Both properties, hyperchaos and dissipativity, have not been previously detected in \textit{C.elegans} locomotion. This is due to the difficulty of estimating the Lyapunov Spectrum of a physical system due to the presence of noise. The precision of our results highlights the success of dynamic diffraction as a method of studying microorganism movement with minimal noise, hinting towards its application for other species. In addition, the observations made for the \textit{C.elegans} system are key for designing a mathematical or computational model of the nematodes' neuromuscular system in the future.

\begin{acknowledgments}
The authors thank Corey Lynn Murphey for computational support and Susannah Zhang, Olivia Trader, and Nicolas Hugo Reyes Cardozo for their insights.
This work has been funded by the Vassar Undergraduate Research Symposium Institute (URSI) and the Lucy Maynard Salmon research fund.
\end{acknowledgments}

\nocite{*}
\bibliography{sources}
\end{document}